\documentclass{article}
\usepackage{amsmath,amsfonts,amsthm,amssymb,amscd,color,xcolor,mathrsfs}
\usepackage{amsfonts}
\usepackage{amsmath,amsthm}
\usepackage{hyperref}
\usepackage{latexsym}
\usepackage{array}
\usepackage{amssymb}
\usepackage{enumerate}

\usepackage[notcite,notref]{showkeys}

\usepackage[francais,english]{babel}

\usepackage{color}
\usepackage[latin1]{inputenc}

\DeclareFontFamily{U}{wncy}{}
    \DeclareFontShape{U}{wncy}{m}{n}{<->wncyr10}{}
    \DeclareSymbolFont{mcy}{U}{wncy}{m}{n}
    \DeclareMathSymbol{\Sh}{\mathord}{mcy}{"58}

\theoremstyle{plain}
\newtheorem{theorem}{Theorem}[section]
\newtheorem*{theorem*}{Theorem}

\newtheorem{definition}[theorem]{Definition}

\theoremstyle{remark}

\newtheorem*{lem*}{Lemma}
\newtheorem*{sublem*}{Sublemma}
\newtheorem*{remark*}{Remark}
\newtheorem*{NB*}{NB}

\newcommand{\R}{ \mathbb{R} }

\newcommand{\N}{ \mathbb{N} }

\newcommand{\T}{ \mathbb{T} }

\newcommand{\ve}{\vec\eta}
\newcommand{\vej}{\vec\eta^{\,j}}
\newcommand{\supp}{\mathop{\rm supp}\nolimits}

\newcommand{\hw}{ (H${}_3^{\text{weak}}$)}
\newcommand{\hs}{ (H${}_3^{\text{strong}}$)}

\newcommand{\cD}{ \mathcal{D} }

 \newcommand{\strela}{\rightharpoonup}

\newcommand{\EE}{ {\mathbb E}}

\newcommand{\om}{ \omega }

\renewcommand{\phi}{ \varphi }

\newcommand{\eps}{\varepsilon}

\newcommand{\dist}{{\operatorname{dist}}}

\newcommand{\be}{\begin{equation}}
\newcommand{\ee}{\end{equation}}
\newcommand{\ben}{\begin{equation*}}
\newcommand{\een}{\end{equation*}}

\numberwithin{equation}{section}

\author{Sergei Kuksin\footnote
{{S. Kuksin, Institut de Math\'emathiques de Jussieu-Paris Rive Gauche, CNRS, Universit\'e Paris Diderot, UMR 7586, Sorbonne Paris Cit\'e, F-75013, Paris, France; and
School of Mathematics, Shandong University, Jinan, 250100, PRC; and
Saint Petersburg State University, Universitetskaya nab., St. Petersburg, Russia,}
 e-mail: \href{mailto:Sergei.Kuksin@imj-prg.fr}{sergei.kuksin@imj-prg.fr}}
}

\title
{Ergodicity, mixing and KAM}

\begin{document}

\maketitle

\begin{abstract}
In this note we review recent   progress in the problem of  mixing for a nonlinear PDE of parabolic type, perturbed by a bounded 
random force. 
 \end{abstract}

\section{Introduction}
We are concerned with  evolutionary nonlinear PDEs  under periodic boundary conditions,
 perturbed by  finite-dimensional random force. We write  their solutions $u(t,x)$ as  curves    
 $$
 u^{\omega}(t)= u^{\omega}(t,\cdot) 
  \in H \,
 $$
 where $(H, \|\cdot \|)$ is a certain Hilbert space of functions of $x$ (usually this is a Sobolev space over $L_2$). We are  interested in equations of the form
 \be\label{1}
 \dot u + 
 Lu+B(u) =   \vec\eta(t), \quad u(0)=u_0,
 \qquad u(t)\in H,
 \ee 
 where $L=-\Delta$ (or, more generally, $L=(-\Delta)^a, a>0$)
 is the dissipation,  $B$ is a nonlinearity  (its linear part may be non-zero), and 
 $ \vec\eta(t)=    \vec\eta^{\omega}(t,x)$ is a random force. We assume that eq.~\eqref{1} is well posed if the function $\| \ve(t)\|^2$ is integrable on bounded 
 segments.

 We regard $Lu$ and $  \vec\eta(t)$ as a  perturbation and are 
 the most interested in the case when the unperturbed equation 
\be \label{B}
\dot u + B(u)=0
 \ee
 is a Hamiltonian PDE.  
 The problem of  long time behaviour in  hamiltonian systems \eqref{B} is related to the ergodic hypothesis and  is hopelessly complicated. 
 Instead our  goal is to study the long-time dynamics of  the perturbed eq. \eqref{1}.
 
  Consider  eigen-functions of the operator 
  $L$ (these are  simply  the complex exponents), and label them by natural numbers:  
 $$
 Le_j = \lambda_j e_j,\quad j=1,2,\dots.
 $$
 We  will decompose vectors  $u\in H$ in 
 this basis, $u = \sum_{s=1}^\infty u_s e_s$, and will identify  any  $u\in H$ with  the vector of its Fourier coefficients:
 $$
 u = (u_1, u_2, \dots). 
 $$
 Let us take any  set $M\subset\N$ of  indices $j$, finite or infinite, 
 and consider the 
  subspace 
 $$
 H_M\subset H, \qquad H_M = \text{span} (e_j, \, j\in M).
 $$
 The  random force $\ve$ is assumed to be of the form 
 \be\label{2}
 \vec\eta(t) = \sum_{j \in M} a_j \eta^\omega_j(t)e_j \in H_M\,, \quad \sum a_j^2<\infty,
 \ee
where 
$\eta_j$'s are i.i.d. real random processes.  If $|M|<\infty$, the force $\eta$  is called {\it finite-dimensional}. With this notation eq.~\eqref{1} may be 
written as 
$$
\dot u_j +\lambda_j u_j +B_j(u) = a_j \eta_j^\omega(t),\;\; \;\;j\ge1; 
\qquad \text{$a_j =0 $  if $ j\not\in M$}. 
$$
The  objection  is to show  that a large class of  ``non-degenerate" 
equations  \eqref{1} with finite-dimensional random forces $\eta$  is ``ergodic", more precisely  -- 
  mixing:\\
  Denote by $u(t;u_0)$ a solution of \eqref{1}, equal $u_0$ at $t=0$. It depends on a random parameter 
  $\omega\in \big(\Omega, F,P)$. 
  
 \begin{definition}  Eq. \eqref{1} is called {\it mixing} if in the space $H$  exists a Borel measure $\mu$ such that 
   for any ``reasonable" functional $f: H\to \R$ and for any starting point $u_0$ 
   \be\label{mix}
  \text{the observable \ $ \EE f(u(t;u_0) )$ converges, as $t\to \infty$, to   $ \int_H f(u)\, \mu(du)$}.
   \ee
  This measure $\mu$ is called the   {\it stationary measure} for eq. \eqref{1}. 
   \end{definition}

  Note that \eqref{mix} means that for any  $u_0$, 
   \be\label{mixx}
   \cD (u(t;u_0)) \strela \mu \quad \text{as}\quad t\to\infty,\;\; \text{weakly,}
   \ee
  where $\cD$ signifies distribution of a random variable, 
   and that  
   $$
\dist (\cD u(t,u_{01}), \cD u(t,u_{02})) \to 0 \quad\text{as}\quad t\to \infty, 
\quad\text{for all}\quad u_{01}, u_{02}
$$
(here {\it dist} a distance in the space of measures on $H$ which 
metrises the weak convergence   $\strela$). 
If the convergence \eqref{mixx} is  exponentially fast,  eq. \eqref{1} is called {\it exponentially  mixing}.

  What was known about the mixing in equations \eqref{1}:\\
i) If $H_M =H$, then 
 the mixing is proved for various classes of equations,
  see in~\cite{KS}. 
  
  \noindent 
ii) If the set $M$ is finite,  then what was available is the result of Hairer--Mattingly \cite{HM} 
who proved the mixing for 
the case of  white in time  forces $\ve$. Their proof is 
based on an infinite--dimensional version of the Malliavin calculus and  applies to a rather special class of eq. \eqref{1}, 
which includes the 2d NSE on the torus. In particular, this approach  does not apply if $B(u)$ is a Hamiltonian nonlinearity 
which is a  polynomial of degree $>3$ (this restriction on the degree of nonlinearity 
also remains true for finite-dimensional systems). Even more: for some important equations (B) corresponding equations \eqref{1} with white-noise forces are not
known to be well posed, while equations \eqref{1} with bounded random forces are well posed, and -- as our results imply -- are mixing.
For example, this is the case
for the primitive  equations of atmosphere 
which are principal  equations of meteorology (the stochastic primitive equations are known to be 
well posed only in some weak sense). 
\medskip

Below I present recent result on the mixing in equations \eqref{1} with bounded random forces, recently obtained in \cite{1} and \cite{2}. 
In \cite{2} the approach of the original work \cite{1} is repeated for an easier problem which resulted in a shorter and more accessible text.   
\medskip

{\bf Acknowledgements.} I thank l'Agence Nationale de la Recherche for support through the project   ANR-10-BLAN~0102, and 
the Russian Science Foundation  -- through the grant  18-11-00032.

\section{ Bounded random forces.}
 
 Recall that  the random force $ \vec\eta(t)$ has the form \eqref{2}, 
 where $\eta_1^\omega(t), \eta_2^\omega(t),\dots $ are i.i.d. bounded random processes. To define a suitable class of 
processes $\eta_j$ 
 we use a naive  approach: 
 Let $\{h_1(t), h_2(t), \dots\}$ be a basis of functions on $[0,\infty)$, made by   bounded  functions. We  define
$$
 \eta_j^\omega(t) = \sum_{k=1}^\infty c_k \xi_k^{j\omega} h_k(t),\quad c_k\ne0,
  $$
  where $\{ \xi_k^{j\omega} \}$ are i.i.d., $|\xi_k^{j\omega}| \le1$. So $\eta_j$'s are random series in the basis $\{h_j\}$. 
  For our techniques to apply, we have to impose on the basis  $\{h_j \}$ a restriction. For $j\in\N$ let us denote $J_j =[j-1, j]$. 
  We assume that 
  $$
  \text{for every function $h_l(t)$, its support belongs to some segment $J_j$, $j=j(l)$. 
  }
  $$
   Our favorite example of a base   as above is the  Haar base ``of step 1"   $\{h_{j,l}(t), j,l\ge0\}$. 
   Each function $h_{0,l}$ is a characteristic   function of the segment $[l,l+1]$, while for $j\ge1$ 
   each  $h_{j,l}$ is a ``dipole" of unit $L_2$--norm 
   on the    segment\  $[2^{-j}l, 2^{-j}(l+1)]$:
    \begin{align*}
	h_{j,l}(t)&=\left\{
	\begin{array}{cl}
		0 & \mbox{for $t<l2^{-j}$ or $t\ge (l+1)2^{-j}$},\\[2pt]
		2^{j/2} & \mbox{for $l2^{-j}\le t<\bigl(l+\tfrac12\bigr)2^{-j}$},\\[2pt]
		- 2^{j/2}  & \mbox{for $\bigl(l+\tfrac12\bigr)2^{-j}\le t<(l+1)2^{-j}$}.
	\end{array}
\right.
\end{align*}
This is an orthonormal base of $L_2(0, \infty)$.

   Now consider the random force    
   $\;
   \vec\eta(t,x) = \sum_{j \in M} a_j \eta^\omega_j(t)e_j(x) .
$
We take the processes  $\eta_j(t)$  to be i.i.d. random Haar series:
\be\label{3}
\eta_{1}(t) =
\sum_{k=0}^\infty
c_k\sum_{l=0}^\infty \xi^\om_{k,l}\,  h_{k,l}(t)\,, \quad c_k\ne0.
\ee
Here  $\{\xi^\om_{k,l}, k,l \ge0\}$ are i.i.d. bounded  random variables such that 
 $| \xi_{k,l}|\le1$ a.s. and   $\cD \xi_{k,l} =p(x)\,dx$,
where $p(x)$ is a Lipschitz function, 
$p(0)\ne0$.

It is known that if $c_k\equiv 1$  and  $\{\xi^\om_{k,l}\}$ are independent $N(0,1)$ r.v.,  
then  \eqref{3}  is a white noise.
 We assume that the random process  in eq.~\eqref{1} is
 much smoother than that:  
  the i.i.d. \ r.v. $\{\xi^\om_{k,l}\}$ are bounded and  the process is ``smooth in time":
\be\label{33} 
|c_n|\le C n^{-q}2^{-n/2}, 
\quad  \text{for each\  $n$},
\ee
where $ q>1$. 
Such  processes are called    {\it red noises.} 

Consider any red noise $\eta_1$ as in. \eqref{3}, \eqref{33}, and for $N\in\N$ consider the process
$$
\beta_N^\omega(T) = \frac1{\sqrt N} \int_0^{NT} \eta_1^\omega(t)dt = c_0 \frac1{\sqrt N}  \sum_{l=0}^{[NT]-1} \xi^\omega_{0,l} 
+O \big( \tfrac{NT- [NT]}{\sqrt N}\big). 
$$
Its trajectories are Lipschitz functions of $T$, and by Donsker's invariance principle the process $\beta_N(T)/\sigma$, $\sigma^2= \EE(\xi_{0,0})^2$, 
converges in distribution to the Wiener process. That is, on large time-scales $\int\eta_1$ behaves as a Wiener process. So the red noises 
 are ``smoother  siblings" of the white noise.

In view of \eqref{33} and since $\sum a_j^2<\infty$, the force $\ve(t)$ is bounded in $H$, uniformly in $t$ and $\omega$. 
Since \eqref{1} is a well posed equation of parabolic type, then usually it possesses the following regularity property, 
which is being assumed below: there is a compact set $X\subset H$ such that 
  $$
  \forall\, u_0\in H \;\; \text{there exists} \;\; t(\|u_0\|) \ge0 \;\;\text{such that} \; u(t) \in X \;\; \forall\, t\ge t(\|u_0\|),  \ \forall\,\omega. 
  $$

\section{Shift Operator $S$}
We wish  to pass from  continuous  to  discrete time. To do that 
let  us cut $\R_+$ to the unit segments $J_l, l\ge1$, and   consider the process $\ve$,
restricted to any $J_j$:
 $$
 \ve^{\,j} (t) = \ve(t-(j-1)), \;\; 0\le t\le 1, 
 \qquad  \vej : [0,1]\to H_M.
 $$
 Denote \ \  $E=L_2(0,1;H_M)$.  Then 
 $$
 \text{the law of   $\vej$  is a measure in $\;E$, independent from $j$},
 $$
 and 
 $\ 
  \supp \cD \vej $
  is a compact set in $ E\  $ since the r.v. $\xi_{k,l}$ are bounded and  $\sum a_j^2<\infty$,
   $\sum c_k <\infty $.  \\
   
   \noindent 
{\bf Operator $S$}.  Consider the operator 
$$
S: H\times E\to H, \quad (u_0, \ve^{\,1}) \mapsto u(1); \qquad \text{$u(t)$ -- solution of eq. \eqref{1}}, \ u(0)=u_0.
$$
Then $u(2) = S(u(1), \ve^{\,2}) $, etc. 

Our task is to understand iterations of the operator $S$, i.e. to study the equation
\be\label{S_eq}
u_{k+1} = S(u_k, \ve^{\,k+1}), \quad k\in\N, 
\ee
where  $u_0\in H$ is given. Certainly for $k\in \N$ 
the solution of \eqref{S_eq} after $k$ step equals $u(k;u_0)$.

{\bf Differential  of $S$ in  $\ve$}. For $u\in H, \ve\in E$ consider the linearised in $\ve$ 
map $S$:  
$$
D_{\ve} \,S(u,\ve): E\to H. 
$$
This operator examines how  a solution $u(t)$ at $t=1$ changes when we modify infinitesimally
the force $\ve(t), 0\le t\le1$, keeping $u(0)$ fixed. 
More precisely for any given $u_0\in H$ and $\ve_0\in E$ 
 to calculate $ D_{\ve} S(u_0,\ve_0)(\vec\xi)$,   $\vec\xi\in E$, 
  we do the following:
  find  a solution $u(t)$ of \eqref{1} for $0\le t\le1$ such 
that $u(0)=u_0, \ve=\ve_0$. Linearise eq.\eqref{1} about  this $u(t)$ and add 
to the obtained linear  eq. the r.h.s. $\vec\xi$: 
$$
\dot v +Lv + dB(u(t)) (v(t))= \vec\xi(t), \;\; v(0)=0,
\qquad 0\le t \le1. 
$$
Consider $v(1) \in H$. This is  $  D_{\ve} S(u_0,\ve_0)(\vec\xi)$.

\section{The main theorem }

We require from the shift--operator $S$ the following three properties:\\

\noindent 
 (H${}_1$) (regularity).  a) $S(X\times $ supp$\cD(\vej) )\subset X$ for some compact $X\subset H$, and \\
b) there is a compactly embedded Banach space $ V\Subset H$ such that:  
$$
S:H\times E \to V\; \text{is $C^2$--smooth}.
$$
 (H${}_2$) (stability of 0). If in \eqref{1} $\ve \equiv 0$, then all solutions of \eqref{1} converge to 0 exponentially. 
\smallskip

 (H${}_3$) (approximate linearised controllability). This assumption is a key point. It exists in a strong and weak forms:

 \hs \  For each point $u\in X$ and every $\xi\in E$, $\xi\in \supp\cD(\ve^1)$, the mapping 
$D_{\ve} S(u,\xi) :E\to H$ has dense image in $H$. 

This condition is easy to verify. It holds if $M=\N$ (all modes are excited), 
but it does not hold  if $M$ is a finite set. To work with finite--dimensional  random forces $\ve$  we evoke a weaker condition: \\

 \hw \  For each point $u\in X$ there exists a null-set $\Omega_u$    such that if  $\omega\notin \Omega_u$, then the range of the 
linear operator $D_{\ve} S(u,\ve^\omega)$ is dense in $H$. \\

\noindent 
    FACT (see \cite{1}). 
     If\\
$ \star\;$ eq. \eqref{1} is the  2d NSE, \\ 
 $ \star\;$ or eq. \eqref{1} is the CGL equation    
$$
\dot u- \epsilon \Delta u -i \gamma \Delta u +i|u|^{2p}u=\ve(t,x),\quad \epsilon>0,\; \gamma \ge0, 
 \quad x\in \T^d, 
$$
where \\
a) either $d=2$ and  $p$ is any, \ or \\
 b) $d=3$ and  $p\le2$,  \ or\\
 b) $d$ is any, $p$ is any, $\gamma=0$, 
 
and  the force  $\vec\eta^\omega(t,x)$  is a red noise as above, then:

1) if $M=\N$, then  (H${}_1$)-- \hs \   holds. 

2) if $M$ is a finite set, satisfying some small restrictions,  then (H${}_1$)-- \hw \  hold. 

The hardest is to  check \hw. 
For the 2d NSE similar results were first obtained by Weinan E, Mattingly, Pardoux, Hairer,
next they were 
properly understood by Agrachev--Sarychev, and  developed further  by Shirikyan, Nersesyan 
and others.

\begin{theorem}\label{t_main}
     Equation \eqref{1} is exponentially  mixing if either\\
1)  (H${}_1$)--\hs\  hold,\\
or if \\
2)  (H${}_1$)--\hw\  hold, and  the mapping $S$ is analytic. 
\end{theorem}

The assertion 2) is proved in \cite{1}, and assertion 1) is established in \cite{2}, using the method of \cite{1}.

\section{ How do we prove this? (``Doeblin meets Kolmogorov")}
Let $u(t)\in X$ and  $u'(t)\in X$ be two solutions of \eqref{1} with initial data $u_0$ and $u'_0$. It is not hard to see that in our setting to 
 prove the mixing we should  verify that   
\be\label{4}
\dist (\cD u(t), \cD u'(t)) \to 0 \quad\text{as}\quad t\to \infty,
\ee 
for all $u_0, u_0'$. 
How to establish \eqref{4} ?  \\

\noindent 
{\it Doeblin' coupling}, a.k.a.  {\it the method of two equations}. 
 In $X\times X$ consider the integer-time dynamics $(u_k, v_k)$, $k\ge1$, where 
$$
(u_0, v_0) =(u_0, u'_0) ,
\qquad\quad
(u_k, v_k) =(S(u_{k-1},\ve_k), \, S(v_{k-1},\ve^{\,'}_k),\;\; k\ge1,
$$
with $ \ve^{\,'}_k =  \eta'_k(u_{k-1}, v_{k-1}, \ve_k)$ such that 
\be\label{6}
\cD \ve^{\,'}_k = \cD \ve_k .
\ee
Then for each $k$,  $u_k  = u(k)$ and   
$\cD v_k = \cD u'(k) $. 
If we can choose  $\ve^{\,'}_k, k\ge1$, such that \eqref{6} holds and 
\be\label{*}
\text{$ \| u_k -v_k\| \to0 $ as $ k\to \infty$ a.s., } 
 \ee
then $\dist (\cD u(k), \cD u'(k)) \to 0$ as $k\to\infty$, 
and our goal  \eqref{4} is achieved.

To achieve \eqref{*}, at  Step 1  we wish to choose the kick $\ve^{\,'}_1$, depending on $u_0, v_0$ and $\ve$,
in such a way that 
$\cD \ve^{\,'}_1= \cD \ve_1$,
and    
$$
\text{ $\| u_1 - v_1\| \ $\ is small with hight probability.
}
$$
Then the law of $u_1$ will be ``rather close" to that of $v_1$, and iterating we will get \eqref{*}. 

We have to distinguish two cases:\\
a) $\ \| u_0 -v_0\| \le \delta_0$, \\
${}  \ \  \ \ $ where $\delta_0$ is an additional small parameter;\\
b) $\ \| u_0 -v_0\|  > \delta_0$.

In case b) we choose for  $\ve^{\,'}_1$ an independent copy of  $\ve_1$, 
and 
use the assumption $(H_2)$ (stability of zero) to achieve a) with positive probability, 
 in a few steps.

Now let $\ \| u_0 -v_0\| =\delta \ll1$.  This is the main difficulty. 
Then we choose   
$$
\ve^{\,'}_1 = \Psi_{u_0, v_0}(\ve_1),
$$
where $\Psi$ is an unknown mapping which 
preserves the measure $ \cD \ve_1$,  so $\cD \ve^{\,'}_1= \cD \ve_1$.
 The dream would be to find $\Psi$ such  that   
\be\label{7}
u_1-v_1=
S(u_0, \ve_1) - S(v_0, \Psi_{u_0, v_0}(\ve_1)) =0\qquad\forall\, \ve_1.
\ee
 Then 
 $v_1=u_1$ a.s., and \eqref{4} is achieved. But  this is hardly possible since it is very exceptional 
 that $\cD(S(u_0, \ve^1)) = \cD (S(u_0', \ve^1))$ for $u_0 \ne u_0'$. 
 
 The situation is reminiscent to that treated by Kolmogorov in his celebrated work which initiated the KAM theory. There Kolmogorov considers 
 a perturbation of an integrable Hamiltonian,
 \be\label{KK}
 H^1(p,q) = H_0(p) +\eps h(p,q), \quad \eps\ll1,\;\  (p,q) \in P\times \T^n, 
 \ee
 where $P$ is a domain in $\R^n$. If  exists a canonical transformation $S:   P_1\times \T^n \to  P\times \T^n,$ where $P_1$
 is a large subdomain of $P$, such that
 \be\label{KKK}
  H^1\circ S = H'_0(p),
  \ee
   then the equation with the transformed  Hamiltonian $H^1\circ S$ would be integrable 
 on $ P_1\times \T^n$. Since Poincar\'e it is well known that normally such a transformation $S$ does not exist. So instead of the 
 hopeless equation 
 $$
 H^1\circ S - H'_0(p)=0,\quad S = ? ,
 $$
 Kolmogorov suggested to look for $S$ in the form $S=$id$+\eps S_1$,\footnote{here $S_1$  is a vectorfield, and the expression id$+\eps S$
 should be properly understood.}
 to linearise the equation in $\eps$,
 $$
 (H_0+ \eps h)\circ ( \text{id} +\eps S_1) = H'_0 (p) + \eps h_1(p,q) + O(\eps^2),
 $$
 and to search for an  $S_1$ such that $h_1=O(\eps^2)$. This transformation should be defined for $p$
 from a large subdomain $P_1\subset P$. The term $h_1$ linearly depends on $S$, so the equation
 \be\label{K0}
 h_1(S) =0
 \ee
 is linear in $S$. It is  called {\it  homological equation}, and one looks for its approximate solution with a disparity of
 order $\eps$. 
 If such an $S_1$ exists, then replacing $H^1$ with the transformed 
 Hamiltonian $H^2 =H^1\circ ($id$+\eps S_1)$ we arrive at a Hamiltonian of the form \eqref{KK} but with $\eps$ replaced by
 $C\eps^2$. Then we would iterate the procedure and after infinitely many steps will arrive at a transformation $S$
 which satisfies \eqref{KKK} for all $p$ from a Borel subset of $P$ of large measure.\\
 
 Let us proceed likewise with the impossible 
equation \eqref{7}.
Namely, for $\delta = \|u_0-v_0\| \le \delta_0$  let us re-write the equation, looking for the mapping $\Psi_{u_0, v_0}$ in
the form   $\Psi_{u_0, v_0}=\text{id} +\delta\Phi $
and neglecting in \eqref{7}   terms $\sim\delta^2$. Then eq. \eqref{7} reeds
$$
\delta   [D_{\ve} S(u_0,\ve_1)\Phi(\ve_1) -S^\Delta(u_0,v_0,\ve_1)]+O(\delta^2)=0,
$$
where 
$S^\Delta=\delta^{-1}(S(v_0,\ve_1)- S(u_0,\ve_1))\sim1$. Requiring that the sum of the terms in the square brackets 
vanishes we get the   homological equation: 
\be\label{hom_eq}
 D_{\ve} S(u_0,\ve_1)\Phi  =S^\Delta(u_0,v_0,\ve_1) ,\qquad \Phi = ?
\ee

 -- If \hs\  holds, we can solve the homological equation   
approximately. 

-- If \hw\  holds, we can solve it approximately  for all $\omega$'s outside some bad event  $\Omega^1_{u_0}$ of small measure, like in the
Kolmogorov scheme above, where the homological equation \eqref{K0} may be non-soluble, even approximatively, for $p$ from some small subset of $P$. 

With the solution  $\Phi= \Phi(\ve_1)$ in hands we, as planned, 
choose $\ve^{\,'}_1=\ve_1+\delta \Phi(\ve_1)$. Then 
  
$$
\text{ $\|u_1-v_1\| \ll  \delta$\;\; for $\omega$ outside\;\;   $\Omega^1_{u_0}$.
 }
$$
Note that since the control for the norm of the solution $\Phi$ of \eqref{hom_eq}
is very poor, 
then now, in difference with KAM, {\bf we cannot obtain the quadratic approximation  
$$
\|u_1-v_1\| \ll  \delta^2 =\big(\|u_0-v_0\|\big)^2, 
$$
despite the method we are using is quadratic!} We only can achieve that $\|u_1-v_1\| \le \frac12\delta$. 
But this turns out to be enough  to get the convergence \eqref{4}. \\

 Two main problems appear on the way: 
 
 1) what should we do when    $\omega\notin  \Omega^1_{u_0}$, so we cannot solve  \eqref{hom_eq} 
 approximately?
 
 2) the mapping     $\ve_1\mapsto \ve^{\,'}_1= \Psi(\ve_1)=\ve_1+\delta\Phi(\ve_1)$  does not 
 preserve the measure $\cD(\ve_1)$,  so $\cD(\ve_1) \neq \cD (\ve^{\,'}_1)$. 
 
 The difficulty 1) usually is present in KAM (there we simply throw away the set of bad parameters).
 The second difficulty is  specific for this setting. \\
 What should we do?
 
 {\bf Answer to 1)}. If $\omega\notin  \Omega^1_{u_0}$, we take $\ve^{\,'}_1 = \ve_1$ (the trivial coupling). 
 Then   
 $$
 \|u_1-v_1\| =  \| S(u_0, \ve_1) - S(v_0, \ve_1) \| \le C \| u_0-v_0\| = C\delta,
 $$
 where $C$ is the Lipchitz  constant.  If still $ \|u_1-v_1\| \le \delta_0$, we play the same game. If 
 $ \|u_1-v_1\| >  \delta_0$, we play the game a), i.e., choose $\ve^{\,'}_1$ to be an independent copy of $ \ve_1$.
 
 {\bf Answer to 2)}. Despite $ \cD(\ve^{\,'}_1) \ne  \cD(\ve_1) $, these two laws turn out to be 
  close:   
 $$
 \| \cD(\ve^{\,'}_1) -  \cD(\ve_1) \|_{\text{var}} \le C\delta^a,\quad a>0.
 $$
 This is enough for us:  careful analysis, similar to that in Sections 3.2.2--3.2.3 of \cite{KS}, 
  shows that iterating a) and b) we prove the theorem.

\end{document}